\documentstyle[11pt]{article}      
\begin{document}
\small

\title{Landscape statistics of the p-spin Ising model}

\author{ Viviane M de Oliveira and J F Fontanari \\
Instituto de F\'{\i}sica de S\~ao Carlos\\
Universidade de S\~ao Paulo\\
Caixa Postal 369\\
13560-970 S\~ao Carlos SP\\
Brazil}

\date{}

\maketitle

\centerline{\large{\bf Abstract}}

The statistical properties of the
local optima (metastable states)
of the infinite range Ising spin glass  with $p$-spin 
interactions in the presence of an external magnetic field $h$
are investigated  analytically.
The average number of optima   as well as the typical 
overlap between pairs of identical optima are  calculated 
for general $p$. Similarly to the
thermodynamic order parameter, for $p>2$ and small $h$ the typical 
overlap $q_t$ is a discontinuous
function of the energy.
The size of the jump in $q_t$ increases with $p$ and 
decreases with  $h$,
vanishing at finite values of the magnetic field.

\bigskip

\bigskip

{\bf Short Title:} landscape of the $p$-spin model


\newpage


\section{Introduction}\label{sec:level1}

The emphasis professed by Kauffman on the role of  the topology
of the fitness landscape as a source of order 
in contraposition to natural selection  has arisen considerable
interest in the study  of the statistical  properties
of fitness landscapes \cite{Kauffman}. 
The central issue is the limitation imposed
by the structure of the fitness landscapes on adaptive evolution,
viewed as a local hill-climbing procedure via fitter mutants.
(See \cite{Dennet} for a lucid criticism of these ideas.)
For sake of concreteness, let us consider a population of
asexually reproducing haploid organisms whose genotypes are
described by sequences of $N$ Ising spins ${\bf{s}} =
(s_1, \ldots, s_N)$ with $s_i = \pm 1$. In the discrete space
of the $2^N$ possible  sequences,  evolution
is modelled by an adaptive walk  defined as a
connected walk through a succession of neighboring sequences
(i.e., sequences that differ by a single spin only) each of
which possessing improved fitness \cite{Kauffman}. There are several
questions of interest whose answers may shed  light on the
structure of the landscapes as, for instance,
 the number of fitness
optima in the sequence space and the similarity between these optima,
to mention only those  we will address in this paper.

Most of the analyses have concentrated on the  NK model of 
random epistatic interactions since it possesses
a tunable control parameter that regulates the ruggedness of the
fitness landscape \cite{Kauffman, Levin, Weinberger_1}.
An alternative (and more appealing to the physicists)  
class of fitness functions was proposed
by Amitrano {\em et al.~} \cite{Amitrano}, namely, 
the Ising spin glass with $p$-spin interactions
defined by the random energy function \cite{Derrida,Gross}
\begin{equation}\label{H_p}
{\cal{H}}_p \left ( {\bf s} \right ) = 
- \sum_{1 \leq i_1 <i_2 \ldots < i_p \leq N}
J_{i_1 i_2 \ldots i_p} \, s_{i_1} s_{i_2} \ldots s_{i_p}
- h \, \sum_i s_i
\end{equation}
where the coupling strengths are statistically independent 
random variables with a Gaussian distribution
\begin{equation}\label{prob}
{\cal{P}} \left ( J_{i_1 i_2 \ldots i_p} \right ) =
\sqrt{\frac{N^{p-1}}{\pi p!}} \exp \left [ 
-\frac{ \left( J_{i_1 i_2 \ldots i_p} \right)^2 N^{p-1}}{p!} 
\right ] ,
\end{equation}
and $h$ is the  external magnetic field. In this
context the fitness value ascribed to a sequence or genotype 
${\bf s}$ is the reverse of its energy. Thus the
fitness maxima correspond to the energy minima of (\ref{H_p}).
Henceforth we will refer to the fitness maxima 
or energy minima as  simply optima. 
For $p=1$ or 
$h \rightarrow \infty$  the energy (\ref{H_p}) gives
a single-peaked, smooth correlated landscape, while
the limit $p \rightarrow \infty$ corresponds to the
random energy model of Derrida \cite{Derrida} and yields  an
extremely rugged, uncorrelated landscape. 
The case $p=2$ is the well-known SK model \cite{SK},
which exhibits a large number of highly correlated
local optima \cite{Binder,Mezard}.

For general $p$, little is known  about the statistical features
of the landscape generated by the energy function (\ref{H_p}).
A result  worth mentioning is  that, for $h=0$,
the correlation between values of ${\cal{H}}_p$ for different
configurations is given by \cite{Amitrano, Weinberger_2}
\begin{equation}
\left \langle {\cal{H}}_p \left ( {\bf{s}}^a \right )
{\cal{H}}_p \left ( {\bf{s}}^b \right ) \right \rangle
= \left [ q \left ( {\bf{s}}^a, {\bf{s}}^b  \right ) \right ]^p
\end{equation}
where 
\begin{equation}\label{overlap}
q \left ( {\bf s}^a, {\bf s}^b  \right ) = \frac{1}{N}
\sum_{i=1}^N s_i^a s_i^b
\end{equation}
is the overlap between the two arbitrary states ${\bf{s}}^a$ and
$ {\bf{s}}^b$. Here the average indicated by
$\langle \ldots \rangle$ is taken over the probability distribution
of the couplings (\ref{prob}).
Thus, as mentioned before,
the correlations between energy levels vanish for $p \rightarrow
\infty$. 

The thermodynamics of the $p$-spin Ising model has been
investigated within the replica framework \cite{Gross,
Gardner,Stariolo}. In particular,
for $p=2$ the order parameter function $q(x)$ tends to
zero continuously as the temperature approaches  a critical
value at which the  transition  between the spin glass 
and the high temperature (disordered)  phases takes place
\cite{Binder,Mezard}. For $p \rightarrow \infty$, the system has 
a critical temperature $T_c$ at which it freezes completely
into the ground state: $q(x)$ is a step function with values zero 
and one, and with a break point at $x = T/T_c$ \cite{Gross}.
The situation for finite $p > 2$ is considerably more complicated.
There is a transition from the disordered phase to a partially
frozen phase characterized by a step function $q(x)$ with values
zero and $q_1 < 1$. As the temperature is lowered further, a
second transition occurs, leading to a phase described by
a continuous order parameter function \cite{Gardner,Stariolo}.

The goal of this paper is to investigate the statistical
properties of the fixed points (local or global optima) of adaptive
walks on the fitness landscape defined by equation (\ref{H_p}).
The energy cost of flipping the spin  $s_i$ is
$\delta {\cal{H}}_p = 2 \Delta_i$ where
\begin{equation}
\Delta_i =   \sum_{i_2 <\ldots < i_p}
J_{i i_2 \ldots i_p} \, s_{i} s_{i_2} \ldots s_{i_p} + h s_i .
\end{equation}
is termed the stability of $s_i$. 
Since in an adaptive walk only
flippings or moves that decrease the energy (i.e., increase
the fitness)  are allowed, any state ${\bf s}$ that
satisfies  
\begin{equation}\label{const}
 \Delta_i > 0 ~~~~~\forall i
\end{equation}
is  an optima of the fitness landscape.
Clearly, counting the
number of states that obey (\ref{const}) is equivalent to 
calculating the number of solutions of
the zero-temperature limit of the celebrated TAP equations 
\cite{TAP}. For non-zero temperature, 
the quite involved calculation of the average number of
solutions of the TAP equations has been carried out for $p=2$
\cite{Bray} as well as for general $p$ \cite{Rieger}. However, 
systematic analyses  of the typical energy of the local optima
and of the effects of the external magnetic field have been
undertaken  for the simplest case only,
namely, $p=2$ at zero temperature
\cite{Tanaka,Roberts,Dean}. We  note that in the statistical
mechanics context the local optima are usually termed 
metastable states.
 
In this paper we study at length the effects of 
the magnetic field $h$  on the structure of the local optima
of the $p$-spin energy landscape. More pointedly, we calculate
analytically the average number of local optima with
a fixed energy density $\epsilon$, denoted by $ \langle 
{\cal{N}}(\epsilon) \rangle$.
Although this analysis is quite straighforward, it is justified
since the dependence of that quantity on
$\epsilon$ and $h$ has not been investigated for general $p$.
In fact, we note that  results of extensive numerical simulations 
aimed at measuring $ \langle {\cal{N}} (\epsilon) $
have been reported recently \cite{Mexico}.
More importantly, we calculate the average number of pairs of local optima
with   overlap  $q$ and
fixed energy density $\epsilon$.  This quantity, denoted
by $\langle {\cal{M}} (q,\epsilon) \rangle$, allows us
to determine the typical overlap $q_t$ between pairs of local optima 
with energy density $\epsilon$. Since 
$ \langle {\cal{M}} (q_t,\epsilon) \rangle$ 
is directly related to the second moment of ${\cal{N}} (\epsilon)$,
we can determine the regions in the space of parameters
$(p,\epsilon,h)$  where this random variable is self-averaging.

The remainder of the paper is organized as follows. In
Sec.\ \ref{sec:level2} we derive the formal equation for
the $n$th moment of the random variable  ${\cal{N}} (\epsilon)$.
Then we  use that  result to calculate 
the average number of local optima 
$ \langle {\cal{N}} (\epsilon) \rangle$
in Sec.\ \ref{sec:level3},  
and  the average number of pairs
of local optima $ \langle {\cal{M}} (q,\epsilon) \rangle$
in Sec.\ \ref{sec:level4}. Finally, some concluding remarks are 
presented in Sec.\ \ref{sec:level5}.
%
%
\section{The formalism}\label{sec:level2}
The number of local optima  ${\cal{N}}(\epsilon) $ with 
fixed energy density $\epsilon$ can be calculated by introducing
the quantity $Y_{\bf s}$ defined by
\begin{equation}\label{Y}
Y_{\bf s}  = \left \{ \begin{array}{ll}
            1 & \mbox{if 
      $\epsilon N =  {\cal{H}}_p \left ( {\bf s} \right ) $ and
            $\Delta_i > 0 ~~\forall i$} \\
            0 & \mbox{otherwise}
            \end{array}
            \right. ,
\end{equation}
so that
\begin{equation}
{\cal{N}}(\epsilon) = \mbox{Tr}_{\bf s} Y_{\bf s},
\end{equation}
where $\mbox{Tr}_{\bf s}$ denotes the summation over the 
$2^N$  states of the system.
We are interested in the evaluation of the moment 
$\langle \left [{\cal{N}}(\epsilon)\right]^n \rangle$ for $n =1,2$, which
can be written as
\begin{eqnarray}
\langle \left [{\cal{N}}(\epsilon) \right ]^n \rangle & = &
\left \langle \prod_{a=1}^n \mbox{Tr}_{{\bf s}^a} Y_{{\bf s}^a}
\right \rangle \nonumber \\
& = &  \mbox{Tr}_{{\bf s}^1} \ldots \mbox{Tr}_{{\bf s}^n}
      {\cal W} \left ( Y_{{\bf s}^1} = 1,\ldots, Y_{{\bf s}^n} = 1
      \right )
\end{eqnarray}
where       
${\cal W} \left ( Y_{{\bf s}^1} = 1,\ldots, Y_{{\bf s}^n} = 1 \right )$
is the joint probability that the $n$ random variables
$ Y_{{\bf s}^1},\ldots, Y_{{\bf s}^n}$ assume the value $1$. Using
the definition
\begin{equation}
{\cal{W}}\left ( Y_{{\bf s}^1} = 1,\ldots, Y_{{\bf s}^n} = 1 \right ) = 
\left \langle \prod_{a=1}^n
\delta \left [ \epsilon - {\cal{H}}_p \left ( {\bf s}^a \right )/N
\right] \prod_i \Theta \left ( \Delta_i^a \right ) \right \rangle ,
\end{equation}
the equation for the $n$th moment becomes
\begin{equation}\label{N_fin}
\langle \left [{\cal{N}}(\epsilon)\right ]^n \rangle = 
\left \langle \prod_{a=1}^n \mbox{Tr}_{{\bf s}^a}
\delta \left [ \epsilon N - {\cal{H}}_p \left ( {\bf s}^a \right )
\right] \prod_i \Theta \left ( \Delta_i^a \right ) \right \rangle .
\end{equation}
where $\Theta (x) = 1$ if $x > 0$ and $0$ otherwise. We have presented
the derivation of equation (\ref{N_fin}) in detail  because some
authors have written the random variable ${\cal{N}}(\epsilon)$
in terms of the delta function directly \cite{Roberts,Dean}.
Clearly, this procedure is correct only for the moments of
${\cal{N}}(\epsilon)$ as shown above.

In the next two sections we concentrate on the explicit evaluation
of equation (\ref{N_fin}) for $n=1$ and $2$. To facilitate those
calculations, we   express
the energy $ {\cal{H}}_p \left ( {\bf s} \right )$ 
in terms  of the stabilities $\Delta_i$,
\begin{equation}\label{aid}
{\cal{H}}_p \left ( {\bf s} \right ) =
-\frac{1}{p} \sum_i \left ( \Delta_i + h(p-1)  s_i \right ) ,
\end{equation}
so that the dependence on the couplings in equation (\ref{N_fin})
appears only through the stabilities $\Delta_i$.

\section{Average number of optima}\label{sec:level3}
Using the integral representation of the delta function
and the auxiliary relation
(\ref{aid}) we can write the first moment of ${\cal{N}}(\epsilon)$
as
\begin{eqnarray}
\langle {\cal{N}}(\epsilon) \rangle & = & 
\int_{-\infty}^\infty  \frac{d\tilde{\epsilon}}{2 \pi} 
\exp \left ({\bf i} N \epsilon \tilde{\epsilon} \right )  
\prod_i \int_{-\infty}^\infty  \frac{d\Delta_i d \tilde{\Delta}_i}{2 \pi} 
\Theta \left ( \Delta_i \right) 
\exp \left ( {\bf i} \Delta_i \tilde{\Delta}_i \right ) \nonumber \\ 
&  & \times \mbox{Tr}_{{\bf s}}
\exp \left [-{\bf i} h \sum_i \tilde{\Delta}_i s_i 
+\frac{{\bf i}}{p} \tilde{\epsilon} 
\sum_i \left( \Delta_i +h(p-1) s_i \right )
\right ] \nonumber \\
&  & \times  \left \langle \exp \left (
      - {\bf i} \sum_i \tilde{\Delta}_i 
      \sum_{i_2 < \ldots < i_p} J_{i i_2 \ldots i_p} 
       s_{i} s_{i_2} \ldots s_{i_p}  \right ) \right \rangle .
\end{eqnarray}
The  average over the couplings can be easily carried out
using the identity 
\begin{equation}\label{id_1}
\sum_i \tilde{\Delta}_i  \sum_{i_2 < \ldots < i_p} J_{i i_2 \ldots i_p} 
       s_{i} s_{i_2} \ldots s_{i_p} =
\sum_{i_1 < \ldots < i_p} \left ( \sum_{k=1}^p
\tilde{\Delta}_{i_k} \right )
J_{i_1  \ldots i_p}  s_{i_1} \ldots s_{i_p} 
\end{equation}
and  yields, in the limit $N \rightarrow \infty$,
\begin{eqnarray}\label{av_fin}
\left \langle \ldots \right \rangle &  = &
 \exp \left [ - \frac{p!}{4 N^{p-1}}
\sum_{i_1 < \ldots < i_p} \left ( \sum_{k=1}^p \tilde{\Delta}_{i_k} 
\right )^2 \right ] \nonumber \\
& = & \exp \left [ 
-\frac{p}{4} \sum_i \left ( \tilde{\Delta}_i \right )^2
-\frac{p(p-1)}{4N}  \left ( \sum_i \tilde{\Delta}_i \right )^2
\right ] .
\end{eqnarray}
The remaining calculations are straightforward: a 
Gaussian transformation allows us to  decouple the sites
in (\ref{av_fin}), so that  the integrals
over $\Delta_i$ and $\tilde{\Delta}_i$ as well as the trace over
the spins
can be readily performed.
As usual, we conclude the calculation by carrying out a 
saddle-point integration over two appropriately rescaled
saddle-point parameters.
The final result for 
the exponent $f$ in 
$\langle {\cal{N}}(\epsilon) \rangle = \mbox{e}^{Nf}$
is 
\begin{eqnarray}\label{f}
f  & = & 
 \frac{\epsilon \nu}{\sqrt{p}} - \frac{1}{p-1} \left( \mu^2
- \mu \nu  +\frac{\nu^2}{4p} \right ) - \ln 2 \nonumber \\
& & + \ln \left [ \mbox{e}^{ \bar{h}  \nu }
\mbox{erfc} \left (
-\mu - \bar{ h} \right ) + \mbox{e}^{- \bar{h}  \nu}
 \mbox{erfc} \left (
-\mu + \bar{h} \right ) \right ] ,
\end{eqnarray}
where 
\begin{equation}\label{h}
 \bar{h} = \frac{h}{\sqrt{p}} .
\end{equation} 
Here the saddle-point parameters $\nu$ and $\mu$ are obtained by 
solving the equations $\partial f/\partial\nu = 0$
and $\partial f/\partial \mu = 0$ simultaneously.
In figure 1 we present the exponent $f$ as a function of $\epsilon$
for $p=2$ and several values of $h$. For sake of clarity we present
only positive values of $f$. The decrease in 
the number of local optima as $h$ increases indicates that
the landscape becomes smoother, as expected. The results
for $p > 2$ are qualitatively similar, except that the peaks
are higher and slightly broader.
Two values of the 
energy density  are particularly important, namely, the value
at which $f$ reaches its  maximum value $f_t$, denoted by
$\epsilon_t$,
and the lowest value of $ \epsilon $ for  which $f$ vanishes,
denoted by $\epsilon_0$. While $\epsilon_t$ gives the typical
value of the energy density of the local optima, $\epsilon_0$ 
gives a lower bound to the ground state energy density of the spin
model defined by the hamiltonian (\ref{H_p}) \cite{Tanaka}.
In figures 2 an 3 we present $\epsilon_t$ and $f_t$, respectively,
as a function of $h$ for several values of $p$. These quantities
are easily obtained by setting $\nu = 0$ in equation (\ref{f}).
The single saddle-point equation $\partial f/\partial\mu = 0$
possesses either one root (for either small or large values of $h$) or 
three roots (for intermediate values of $h$).
The discontinuity in $\epsilon_t$ that can be observed in 
figure 2 for $p \geq 7$  is due to the simultaneous
disappearance of two of those roots.
For $p \rightarrow \infty$ and finite $h$ we find
$\epsilon_t \rightarrow \langle {\cal{H}}_p \rangle = 0$, 
signaling thus the emergence
of the so-called complexity catastrophe, i.e., the  energy density
of typical local optima  equals the expected energy of a randomly
chosen state \cite{Kauffman}.
We note that
$\langle {\cal{N}}(\epsilon_t) \rangle = \exp( f_t N)$ 
yields the average number of optima regardless of their energy
values, i.e., the same result is obtained by dropping the
energy constraint in the definition of $Y_{{\bf s}}$ given
in equation (\ref{Y}).
In figure 4 we present $\epsilon_0$ as a function of $h$ for
several values of $p$. 
Clearly, since in the limit
$h \rightarrow \infty$ there is only one optimum, namely,
${\bf s} = {\bf 1}$,  we find
$\epsilon_0 \rightarrow \epsilon_t = -h$. It is important to note
that for $p \rightarrow \infty$, $\epsilon_0 $ tends to a non-zero
limiting value.
This result illustrates the fact that the complexity catastrophe 
phenomenon affects the typical optima only. In fact, the increase of
$p$ has little effect on the ground-state lower bound $\epsilon_0$, 
which for $h=0$ decreases from $-0.791$ for $p=2$ \cite{Bray} to
$-\sqrt{\ln 2} \approx -0.832$ for $p \rightarrow \infty$ \cite{Derrida}.

\section{Average number of pairs of  optima}\label{sec:level4}
We define the
number of pairs of optima with overlap 
$q =-1, -1 + \frac{2}{N},\ldots, 1 $
and energy density  $\epsilon$ as
\begin{equation}\label{M}
{\cal{M}}(q,\epsilon) = \frac{1}{2} 
\mbox{Tr}_{{\bf s}^1} \mbox{Tr}_{{\bf s}^2} Y_{{\bf s}^1} Y_{{\bf s}^2}
\delta \left ( Nq, \sum_i s_i^1 s_i^2 \right )
\end{equation}
where $\delta(m,n)$ is the Kronecker delta and $Y_{{\bf s}}$
is given by equation (\ref{Y}). 
Following the procedure presented in  Sec.\ \ref{sec:level2},
the average of ${\cal{M}}$ over the 
couplings is cast into the form
\begin{equation}\label{M_1}
\langle {\cal{M}}(q,\epsilon) \rangle = \frac{1}{2}
\left \langle  \mbox{Tr}_{{\bf s}^1} \mbox{Tr}_{{\bf s}^2}
\delta \left ( Nq, \sum_i s_i^1 s_i^2 \right ) \prod_{a=1}^2
\delta \left [ \epsilon N - {\cal{H}}_p \left ( {\bf s}^a \right )
\right] \prod_i \Theta \left ( \Delta_i^a \right ) \right \rangle .
\end{equation}
The integral representations of the delta function and
the Kronecker delta allow us to write  this equation as
\begin{eqnarray}
\langle {\cal{M}}(q,\epsilon) \rangle & = & \frac{1}{2} 
\int_{-\pi}^\pi \frac{d\tilde{q}}{2 \pi} 
\exp \left ({\bf i} N q \tilde{q} \right )
\prod_a
\int_{-\infty}^\infty  \frac{d\tilde{\epsilon}^a}{2 \pi} 
\exp \left ({\bf i} N \epsilon^a \tilde{\epsilon}^a \right ) \nonumber \\
& & \times \prod_{ai} \mbox{Tr}_{{\bf s}^a}
\int_{-\infty}^\infty \frac{d\Delta_i^a d \tilde{\Delta}_i^a}{2 \pi} 
\Theta \left ( \Delta_i^a \right) 
\exp \left ( {\bf i} \Delta_i^a \tilde{\Delta}_i^a \right ) \nonumber \\ 
&  & \times \exp \left [-{\bf i} \tilde{q}  \sum_{i}  s_i^1 s_i^2
-{\bf i} h \sum_{ai} \tilde{\Delta}_i^a s_i^a 
+\frac{{\bf i}}{p} \sum_{ai} \tilde{\epsilon}^a  
\left( \Delta_i^a +h(p-1) s_i^a \right )
\right ] \nonumber \\
&  & \times 
       \left \langle \exp \left (
      - {\bf i} \sum_{ai} \tilde{\Delta}_i^a 
      \sum_{i_2 < \ldots < i_p} J_{i i_2 \ldots i_p} 
       s_{i}^a s_{i_2}^a \ldots s_{i_p}^a  \right ) \right \rangle .
\end{eqnarray}
As in the previous section, the average can be performed with
the aid of an identity analogous  to (\ref{id_1}), yielding
\begin{equation}\label{av_2}
\left \langle \ldots \right \rangle   = 
 \exp \left \{ - \frac{p!}{4 N^{p-1}}
\sum_{i_1 < \ldots < i_p} \left [\sum_{a=1}^2 \left 
( \sum_{k=1}^p \tilde{\Delta}_{i_k}^a \right )
s_{i_1}^a \ldots s_{i_p}^a \right ]^2 \right \} .
\end{equation}
After some algebra, 
the argument of this exponential 
is rewritten in the limit $N \rightarrow \infty$  as
\begin{eqnarray}
\left \{ \ldots \right \} &  = & 
-\frac{p}{4} \sum_{a=1}^2 \left [ \sum_i \left ( \tilde{\Delta}_i^a \right )^2
+\frac{p-1}{N}  \left ( \sum_i \tilde{\Delta}_i^a \right )^2 \right ]
-\frac{p \, q^{p-1}}{2} \sum_i \tilde{\Delta}_i^1 \tilde{\Delta}_i^2
s_i^1 s_i^2
\nonumber \\
& &  - \frac{p(p-1)\, q^{p-2}}{2N} 
\left( \sum_i \tilde{\Delta}_i^1 s_i^1 s_i^2 \right )
\left( \sum_i \tilde{\Delta}_i^2 s_i^1 s_i^2 \right ) .
\end{eqnarray}
The next step is to
introduce via delta functions  the auxiliary parameters: 
 $N m_1 = \sum_i
\tilde{\Delta}_i^1 $, $N m_2 = \sum_i
\tilde{\Delta}_i^2 $, 
$N v_{1} = \sum_i \tilde{\Delta}_i^1 s_i^1 s_i^2 $, 
$N v_{2} = \sum_i \tilde{\Delta}_i^2 s_i^1 s_i^2 $, 
and their respective Lagrange multipliers
in order to decouple the variables $s_i^a$ and $\tilde{\Delta}_i^a$
for different sites $i$. Then the integrals over $\Delta_i^a$
and $\tilde{\Delta}_i^a$, and the trace over $s_i^a$ can  be
easily performed. As before, the auxiliary parameters as well as
the Lagrange multipliers $\tilde{q}$ and $\tilde{\epsilon}$ are
integrated out via a saddle-point integration. This part of
the calculation is  straightforward and quite
unilluminating so we do not present any further detail.
To proceed further   we
assume that the symmetry ${\bf s}^1 \leftrightarrow {\bf s}^2$
between the two replicas remains intact, i.e., $m_1 = m_2$
and $v_1 = v_2$. This is a quite sensible
assumption since the  breaking of the replica symmetry that
pervades the thermodynamic calculations \cite{Gross,Gardner,Stariolo}
is very probably a consequence of  the limit where the
number of replicas goes  to  zero. In any event
we  will, conservatively, restrict the forthcoming
analysis to pairs of identical optima only. The final
result for the exponent $g$ in $\langle {\cal{M}}(q,\epsilon) \rangle
= \frac{1}{2} \exp ( g N )$ is  written more simply in 
terms of a new set of
saddle-point parameters that are linear combinations of
those introduced above. We find
\begin{eqnarray}\label{g}
g & = &  \frac{\epsilon \nu}{\sqrt{p}} + q z  - \frac{1}{2(p-1)} \left [
\left ( x+y \right )^2 + q^{2-p} \left ( x- y \right )^2 
 + (1 + q^p) \frac{\nu^2}{4p} \right ] \nonumber \\
& & + \frac{\nu}{2(p-1)} \left [
 \left ( 1 +q \right ) x + \left ( 1-q \right ) y \right ] + 
 \ln \Xi \left ( \nu,x,y,z \right) - \ln 2
\end{eqnarray}
where
\begin{eqnarray}
\Xi & = &  \mbox{e}^{ \nu \bar{h} - z }
\int_{-x - \bar{h}}^\infty Dt \, \mbox{erfc} 
\left [- \frac{x + \bar{h} + q^{p-1} t}{\sqrt{1 - q^{2p-2}}}
\right ] \nonumber \\
& & + \mbox{e}^{ - \nu \bar{h} - z }
\int_{-x + \bar{h}}^\infty Dt \, \mbox{erfc} 
\left [- \frac{x - \bar{h} + q^{p-1} t}{\sqrt{1 - q^{2p-2}}}
\right ] \nonumber \\
& & + \mbox{e}^z
\int_{-y + \bar{h}}^\infty Dt \, \mbox{erfc} 
\left [- \frac{y + \bar{h} - q^{p-1} t}{\sqrt{1 - q^{2p-2}}}
\right ] \nonumber \\
& &  
 + \mbox{e}^z
\int_{-y - \bar{h}}^\infty Dt \, \mbox{erfc} 
\left [- \frac{y - \bar{h} -  q^{p-1} t}{\sqrt{1 - q^{2p-2}}}
\right ] .
\end{eqnarray}
Here $Dt = dt \mbox{e}^{-t^2}/\sqrt{\pi}$ is the Gaussian measure
and $\bar{h}$ is given by (\ref{h}).
The saddle-point parameters $\nu, x,y,z$ must be  determined so as to
maximize $g$. This is achieved by solving
the four coupled saddle-point equations 
$\partial g/\partial \nu =0$, $\partial g/\partial x =0$,
$\partial g/\partial y = 0$, and $\partial g/\partial z =0$.
For $q=1$ we find $y = 0$ and hence $g = f$, as expected.
Furthermore, for $q=0$ and $h=0$ we find $x=y$ and $z=0$ so that
$g = 2 f$.  
Once 
$\langle {\cal{M}}(q,\epsilon) \rangle $
is known, the second moment of ${\cal{N}}(\epsilon)$ can be
calculated using the identity
\begin{eqnarray}
\sum_q \langle {\cal{M}}(q,\epsilon) \rangle & = &  \frac{1}{2}
 \langle \left [ {\cal{N}} (\epsilon) \right ]^2 \rangle 
 \nonumber \\
 & \approx & \langle {\cal{M}}(q_t,\epsilon) \rangle ,
\end{eqnarray}
since  the sum is dominated by the 
overlap  $q = q_t$ that maximizes equation (\ref{g})
in the limit $N \rightarrow \infty$. Hence  we
have $ \langle \left [{\cal{N}} (\epsilon) \right ]^2 \rangle 
= \exp \left ( f^{(2)} N \right ) $ with $f^{(2)}$ given by (\ref{g})
calculated at $q_t$. 
Thus  for $h=0$ the variance of the
random variable ${\cal{N}}(\epsilon)$  vanishes in the thermodynamic
limit, provided that $q_t = 0$. We note that although $q=0$
is always a point of maximum of $g$ for $h=0$, that maximum may not be
the global one and, in that case, $q_t \neq 0$.

For fixed $q$, the dependence of $g$ on $\epsilon$  is similar to that 
shown in figure 1.  Likewise, the maximum of $g$ with respect
to $\epsilon$, denoted by $g_t$, 
is determined by setting $\nu = 0$. In figure
5 we show this maximum as a function of $q$ for $p=7$ and several
values of $h$. The quantity $ \frac{1}{2}\exp \left ( g_t N \right )$
can be viewed as the number of pairs of identical optima
(in the sense that  their energies and saddle-point parameters
are identical)
with overlap $q$, regardless of the specific
value of their energies. 
For $h$ not too large there appears a minimum 
for $q \approx 1$, 
indicating that around a  typical optimum there is a region
where other optima are rarer. The picture that emerges 
is one of clusters of many optima surrounded
by comparatively smoother valleys.
The typical energy $\epsilon_t$ of these optima is shown in figure 6 
as a function of the overlap $q$. 
The typical overlap  $q_t$  between the optima increases
from zero at $h=0$ to one in the limit $h \rightarrow \infty$ since, 
as expected, the external magnetic field
induces correlations between the optima. This is shown in figure 7, 
where we present
$q_t$ as a function of $h$ for several values of $p$. 
The discontinuity that appears for 
$ p \geq 7$ is caused by the
competition between the two maxima shown in figure 5.

Next we consider the dependence of the typical overlap 
between identical optima on their energies. This
analysis is more involved since, besides the four saddle-point
equations, we have to solve the equation $\partial g/\partial q =0$
too.
In figures 8 and 9  we show $q_t$ as a function of $\epsilon$
for $p=2$ and $p=3$, respectively,  and several values of
the external magnetic field. For $h=0$, in both cases we find
$q_t =0$ up to a certain value of the energy density
($\epsilon = -0.672$ for $p=2$ and $\epsilon = -0.792$ for $p=3$).
Thus, as mentioned before,
${\cal{N}}(\epsilon)$ is self-averaging in this regime.
Our results for $p=2$  are remarkably similar to those found
in the replica calculation of the quenched average
$ \langle \ln {\cal{N}}(\epsilon) \rangle$, with the typical
overlap $q_t$ replaced by the saddle-point parameter
$\hat{q} = \langle \langle s_i \rangle^2_\epsilon \rangle$
\cite{Roberts}. Here
$\langle \ldots \rangle_\epsilon$ means an average over
optima with energy density $\epsilon$. In particular,
$\hat{q}$ vanishes for $\epsilon > - 0.672$, indicating thus
that ${\cal{N}}(\epsilon)$ is self-averaging in this regime,
in agreement with our results.
However, while for $p=2$, $q_t$ increases continuously from zero,
for $p = 3$ there is a discontinuity at $\epsilon = - 0.792$.
The same phenomenon is observed for $p > 3$, with
the size of the jump in $q_t$ increasing with $p$. 
This finding is reminiscent of the jump in the order
parameter found in the thermodynamic calculations for
$p >2$ \cite{Gardner,Stariolo}. The  discontinuity
in $q_t$ can be understood by studying the dependence of
the exponent $g$ on the overlap $q$ for $p=3$ and $h=0$, shown
in figure 10. 
Since the typical overlap is associated to the global
maximum of $g$, the competition between
the maximum at  $q=0$
and the maximum at $q > 0$ originates the jump in $q_t$,
which takes place  at the energy density  where the two maxima
have precisely the same height.
The situation for non-zero $h$ is more complicated.
The correlations induced by the magnetic field destroy the
region of self-averageness of ${\cal{N}}(\epsilon)$. Interestingly,
for a given $h > 0$ there is value of the energy density for which 
the typical overlap is minimal. For $p=3$ 
the effect of the magnetic field
is to decrease the size of the jump in $q_t$
till it disappears altogether for $h \approx 0.29$. 
The results for $p > 3$ are qualitatively similar to those
for $p=3$. We mention only that the larger $p$, the larger
the value of $\epsilon$ at which the discontinuity occurs, and
the larger the value of $h$ at which it disappears.
 Unfortunately, the enormous
difficulty of solving the system of five coupled equations
prevents a more systematic analysis of these 
discontinuities.

\section{Conclusion}\label{sec:level5}

The analytical investigation of the statistical structure of the
energy landscape of the $p$-spin Ising model presented in this
paper is of interest  from the viewpoint
of the traditional statistical mechanics
of disordered systems 
\cite{Gross, Tanaka,Roberts, Dean} as well as from the perspective
of the study of adaptive walks in rugged fitness landscapes
\cite{Amitrano, Weinberger_2,Mexico}.
Besides extending the calculation of the average number of optima 
to general $p$ and non-zero magnetic field,
we have focused on the  characterization of the typical
overlap $q_t$ between pairs of identical optima. Interestingly, the
dependence of $q_t$  on the energy density $\epsilon$
is reminiscent of the  dependence of the thermodynamic 
order parameter on the temperature $T$ 
\cite{Gardner, Stariolo}. We must
note, however, that there is
no relation between $T$ and $\epsilon$ since 
$\ln {\cal{N}}(\epsilon)$ is not the entropy of the spin system.
The quite complex effect of the magnetic field on the statistical
properties of the energy optima motivates
a more detailed study of 
the thermodynamics of the $p$-spin model for non-zero $h$. 
In fact, even  the unambitious analysis of the first moment 
$\langle {\cal{N}}(\epsilon) \rangle$ has unveiled
an interesting interplay between $h$ and $p$ that lead to a
discontinuity in the typical energy density of the optima. 
Moreover, we have found that the magnetic field decreases 
the size of the jump in  the typical overlap $q_t$ that occurs
for $p>2$.
It would be interesting to investigate
whether a  similar effect occurs for the thermodynamic order parameter
as well, which might lead, eventually, to a 
continuous phase transition.

To conclude, we must mention that the calculations presented in
this paper are free of all the mathematical subtleties that
permeate the replica analyses of the infinite range 
Ising spin glass \cite{Binder, Mezard}. Thus our results 
present a reliable account of the statistical
properties  of 
the $p$-spin  energy landscape which, though  may have 
little relevance
to the  thermodynamics of the model, are of considerable
interest to the characterization of the fixed points
(metastable states)
of adaptive walks (zero-temperature Monte Carlo dynamics)
on that landscape.

%

\bigskip

{\bf Acknowledgments}
This work was  supported in part by Conselho Nacional de
Desenvolvimento Cient\'{\i}fico e Tecnol\'ogico (CNPq). 
VMO holds a {\small{FAPESP}} fellowship.

\bigskip


\newpage
%


\newpage

\section*{Figure captions}
\bigskip

\parindent=0pt

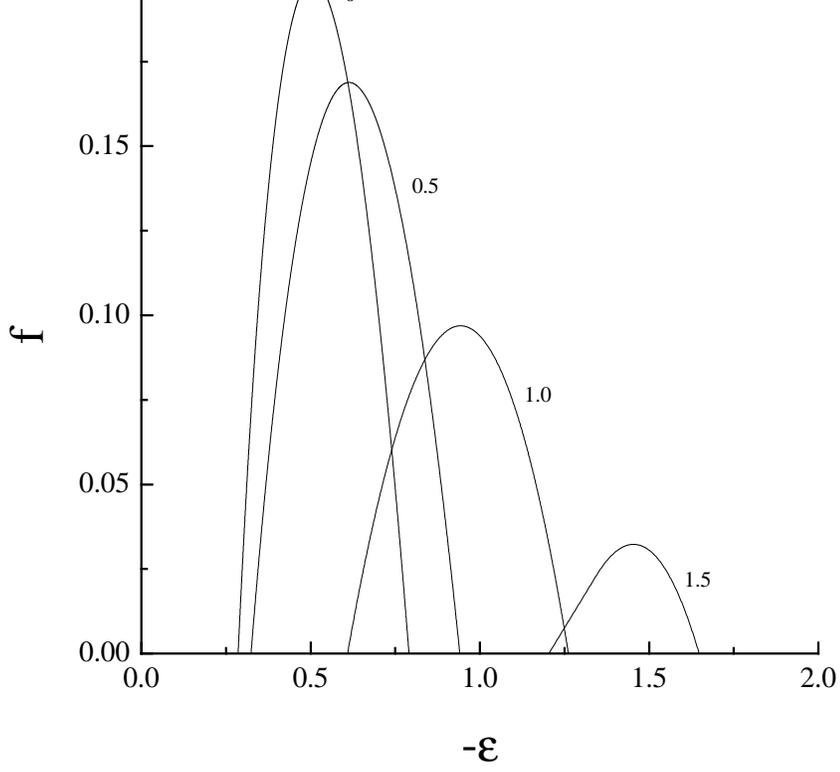
{\bf Fig. 1} The exponent $f$ in $\langle {\cal {N}} (\epsilon) \rangle
= \mbox{e}^{f N}$ as a function of the energy density $\epsilon$
for $p=2$ and $h =0$, $0.5$, $1.0$, and $1.5$.

\bigskip

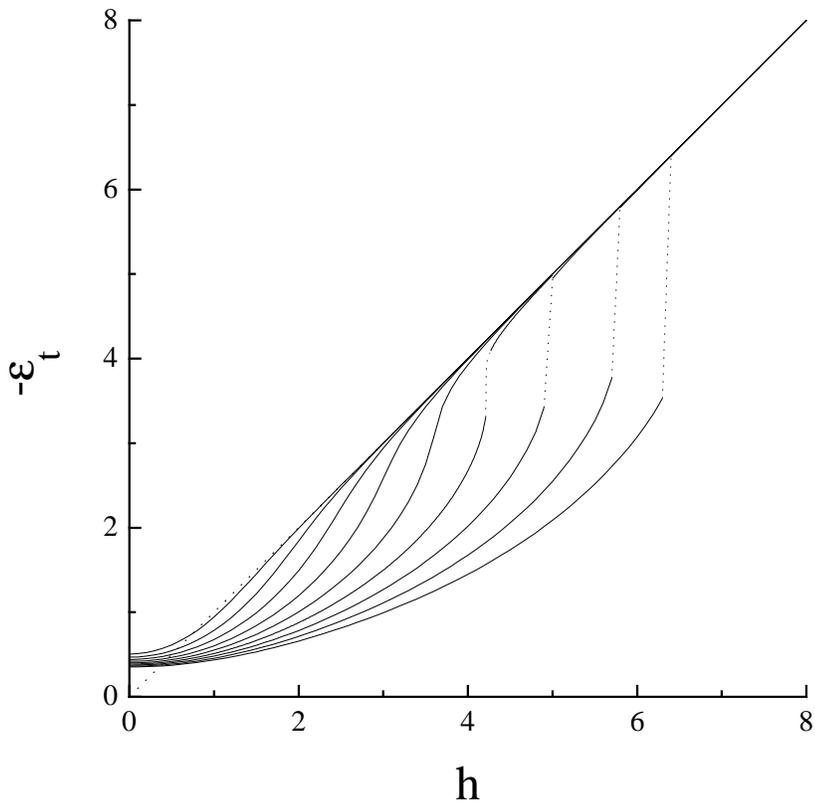
{\bf Fig. 2} The typical energy density $\epsilon_t$ of the 
local optima as a function of $h$ for (from top to bottom)
 $p=2$ to $p=10$. For $ p \rightarrow \infty$ we find
$\epsilon_t \rightarrow 0$. The dashed straight line is 
$\epsilon_t = - h$.

\bigskip

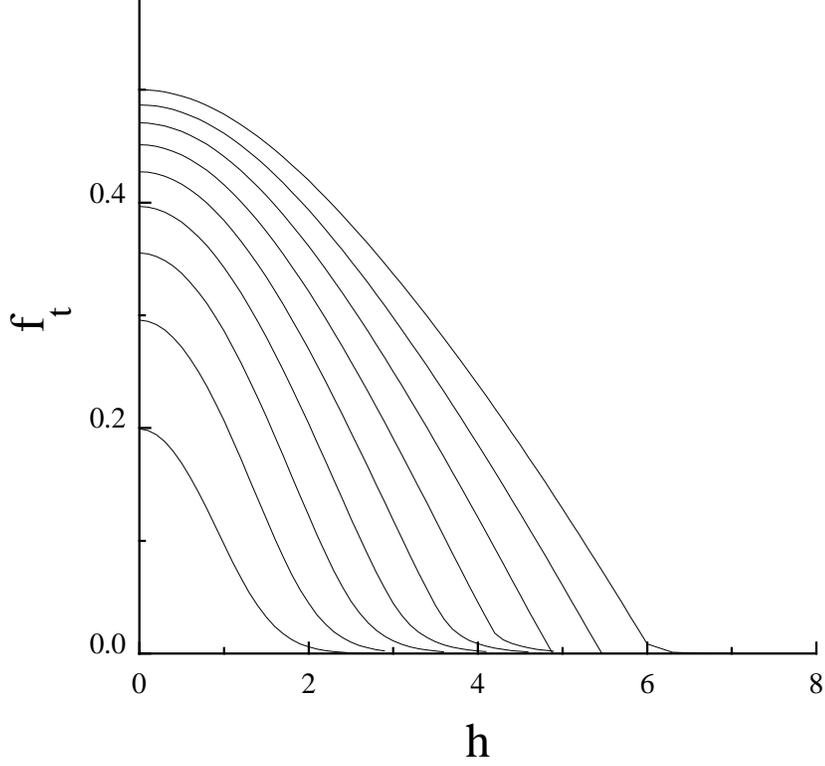
{\bf Fig. 3} The exponent $f_t$ in the expression for the average number
of optima
$\langle {\cal {N}} (\epsilon_t) \rangle
= \mbox{e}^{f_t N}$
as a function of $h$  for (from  bottom to top)
 $p=2$ to $p=10$. 
For $ p \rightarrow \infty$ we find
$f_t \rightarrow \ln 2$.

\bigskip

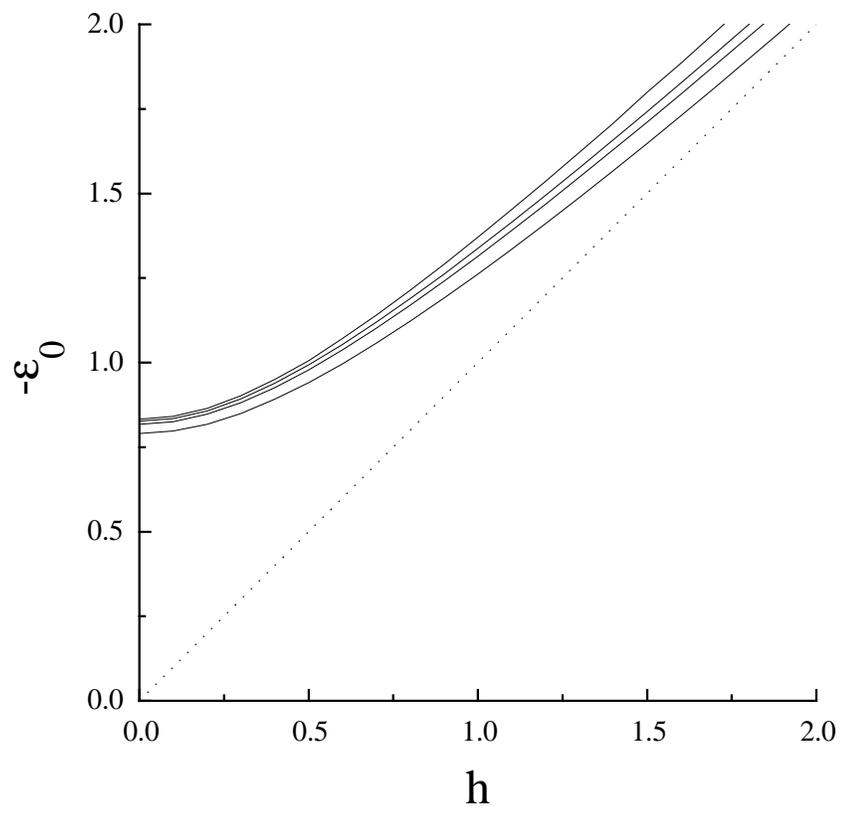
{\bf Fig. 4} The lower bound $\epsilon_0$ to the ground state 
energy density as a function of $h$ for (from bottom to top)
 $p=2$, $3$, $4$, and $\infty$. The dashed straight line is 
$\epsilon_0 = - h$.

\bigskip

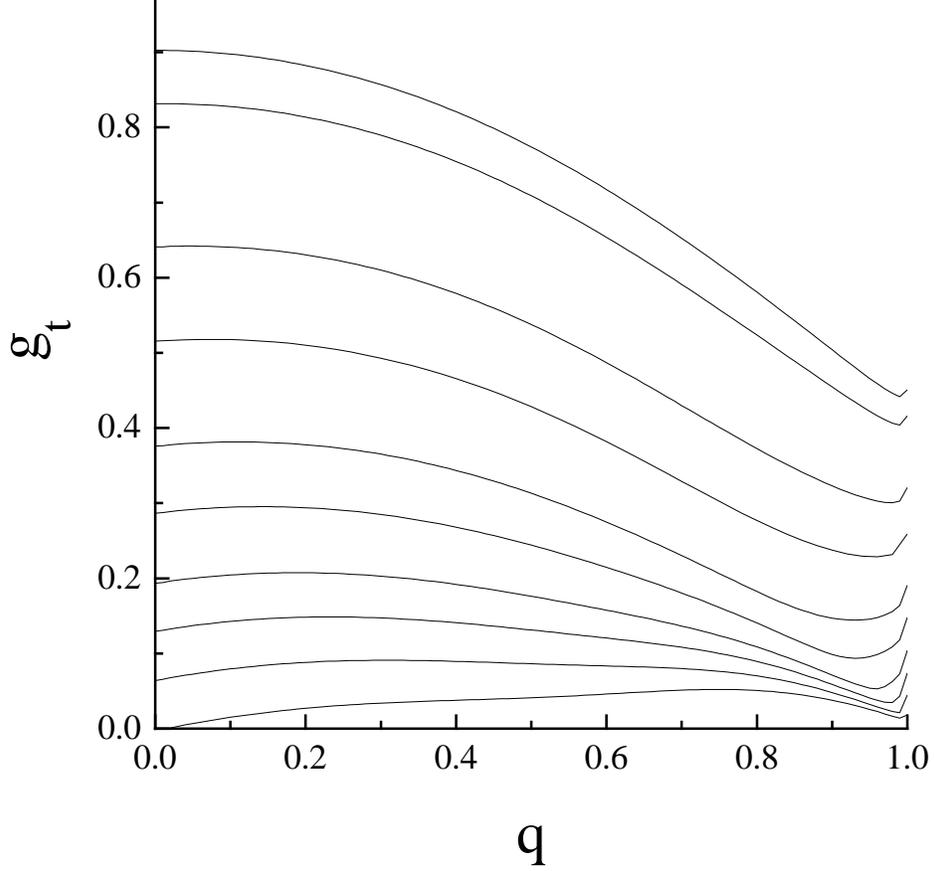
{\bf Fig. 5} The exponent $g_t$ in the expression for the average number
of pairs of identical optima 
$\langle {\cal {M}} (\epsilon_t, q) \rangle
= \mbox{e}^{g_t N}$
as a function of $q$ for $p=7$ and (from top to bottom)
$h=0$, $1$, $2$, $2.5$, $3$, $3.3$, $3.6$, $3.8$, $4$ and  $4.2$.

\bigskip

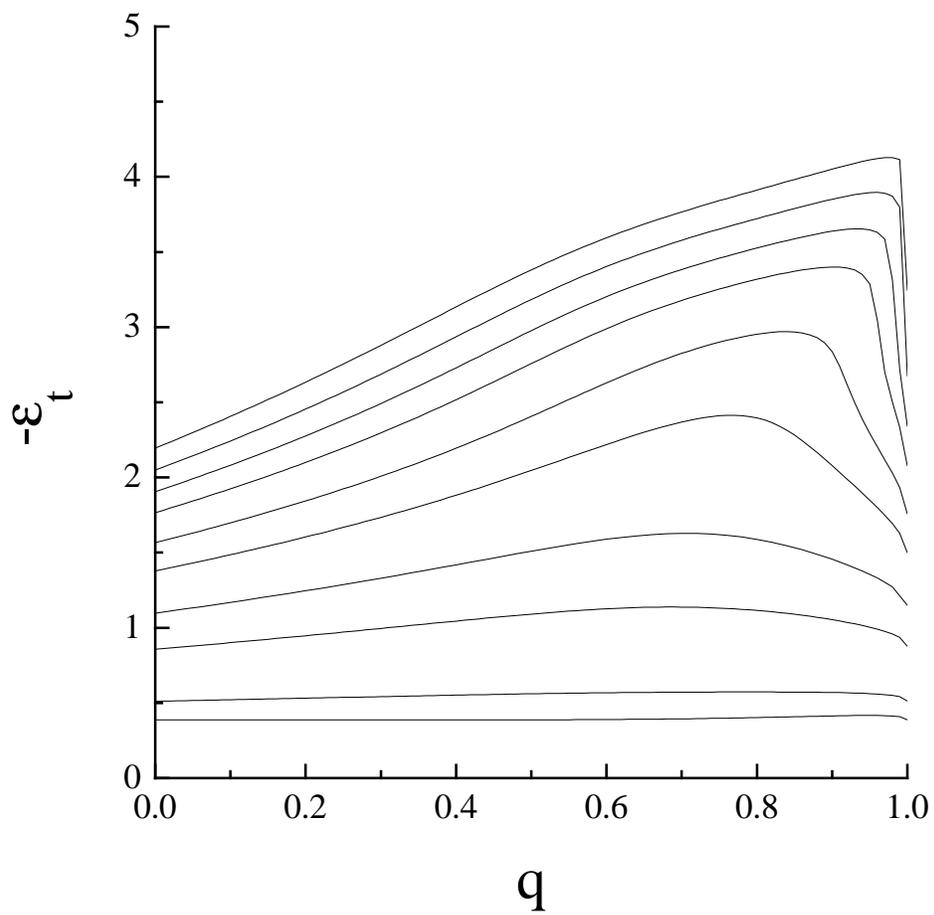
{\bf Fig. 6} The typical value of the energy density of
a pair of identical optima
as a function of the overlap $q$ for $p=7$ and (from bottom
to top) 
$h=0$, $1$, $2$, $2.5$, $3$, $3.3$, $3.6$, $3.8$, $4$ and  $4.2$.

\bigskip

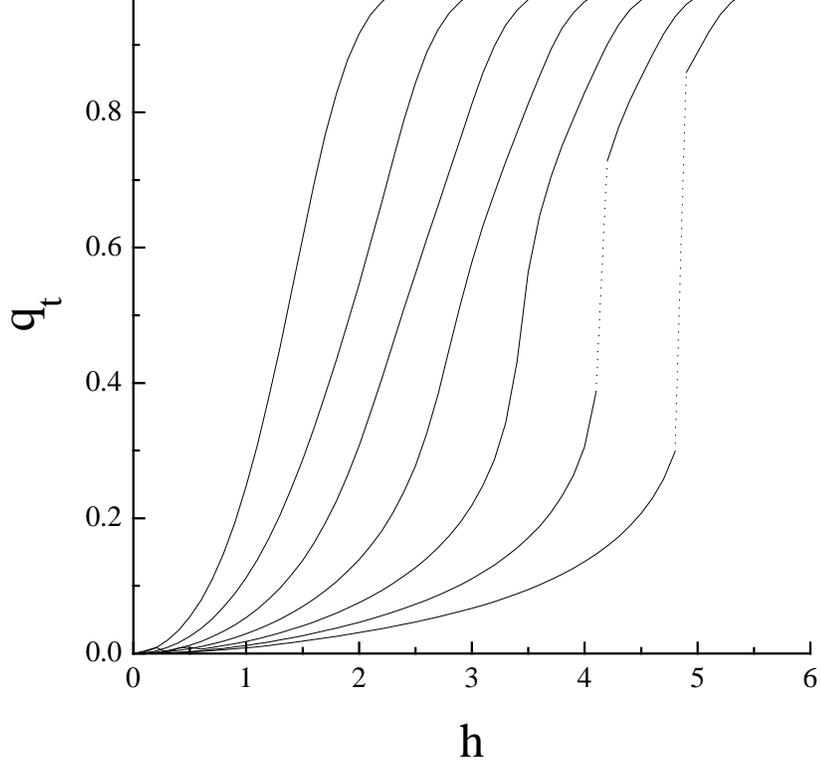
{\bf Fig. 7} The typical value of the overlap between
pair of identical optima
as a function of $h$ for  (from left to right)
$p=2$ to $p=8$.

\bigskip

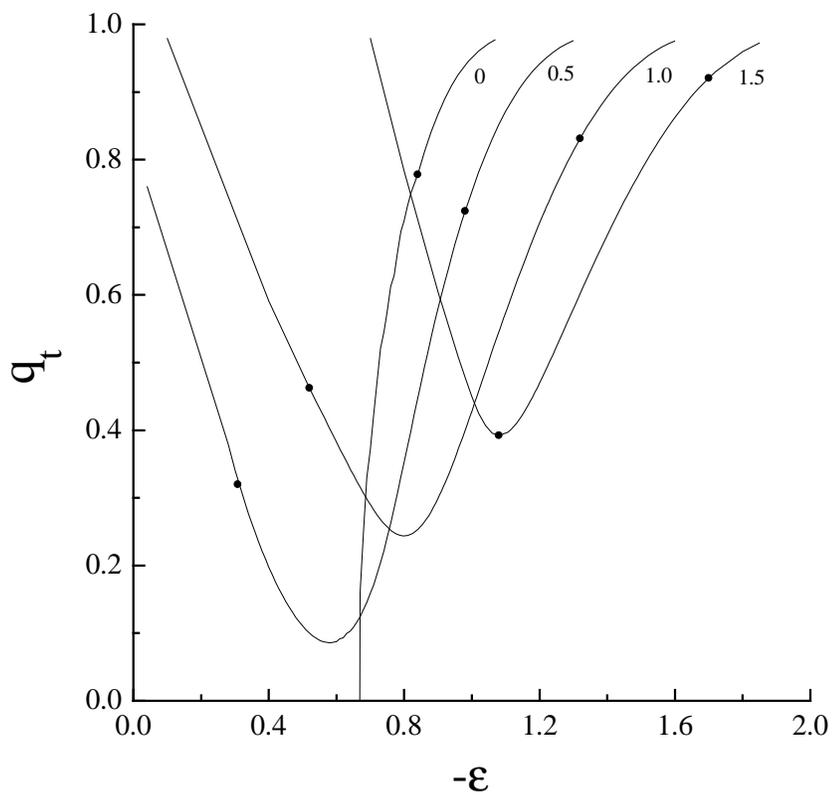
{\bf Fig. 8} The typical value of the overlap between
pair of identical optima as a function of their energy density
for $p=2$ and $h = 0$, $0.5$, $1.0$, and $1.5$. The marked points
correspond to $f^{(2)} = 0$.

\bigskip

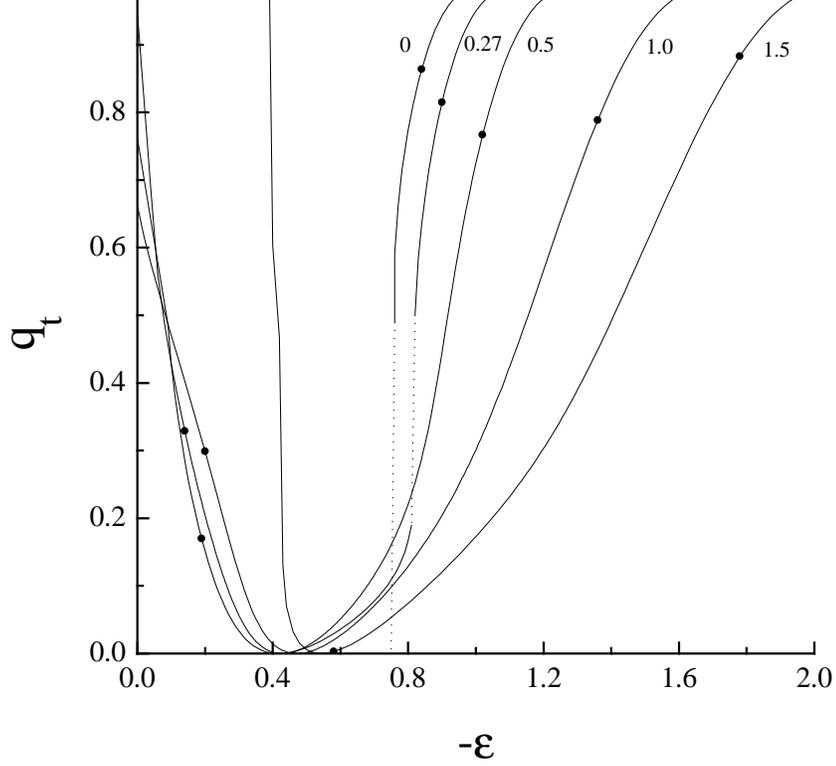
{\bf Fig. 9} Same as figure 8 but for $p=3$,
 and $h = 0$, $0.27$, $0.5$, $1.0$, and $1.5$.

\bigskip

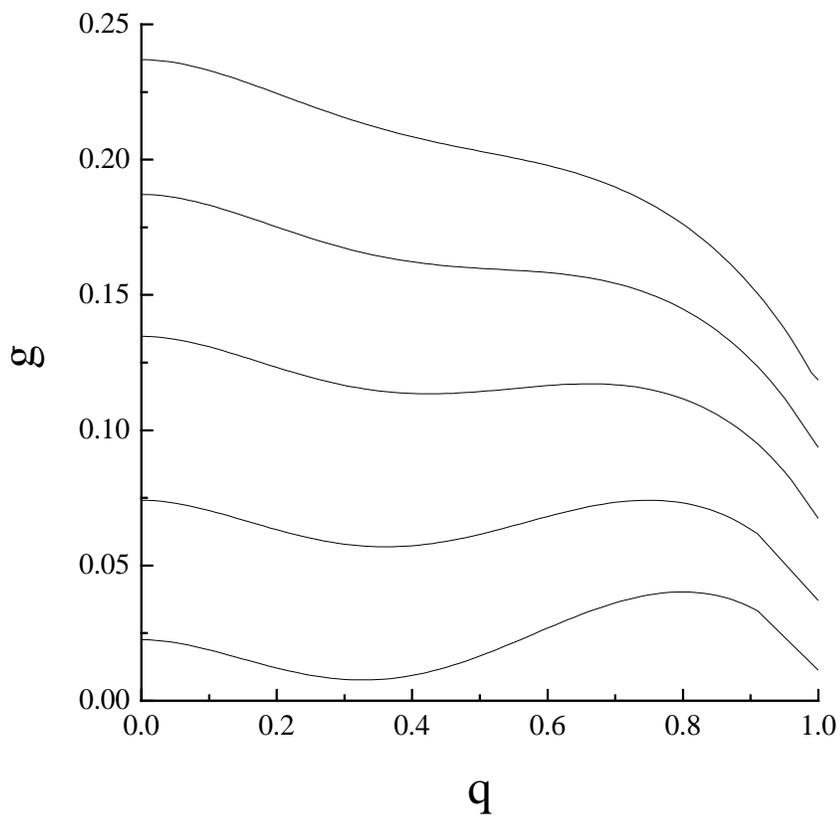
{\bf Fig. 10} The exponent $g$ in the expression for the average number
of pairs of identical optima 
$\langle {\cal {M}} (\epsilon_, q) \rangle
= \mbox{e}^{g N}$
as a function of $q$ for $p=3$, $h=0$, and (from top to bottom)
$\epsilon =-0.73$, $-0.75$, $-0.77$, $-0.79$, and  $-0.81$.

\end{document}